\documentclass[reprint,superscriptaddress,nofootinbib,amsmath,amssymb,aps]{revtex4-2}
\usepackage{graphicx}
\usepackage{dcolumn}
\usepackage{bm}
\usepackage{hyperref}
\hypersetup{colorlinks,citecolor=blue,linkcolor=blue,urlcolor=blue,}

\begin{document}

\title{Constraints on Lorentz-invariance violation in the
neutrino sector from the ultrahigh-energy event KM3-230213A}
\author{Yu-Ming Yang}
\email{yangyuming@ihep.ac.cn}
 \affiliation{%
 Key Laboratory of Particle Astrophysics, Institute of High Energy Physics, Chinese Academy of Sciences, Beijing 100049, China}
\affiliation{
 School of Physical Sciences, University of Chinese Academy of Sciences, Beijing 100049, China 
}%
\author{Xing-Jian Lv}
\email{lvxj@ihep.ac.cn}
 \affiliation{%
 Key Laboratory of Particle Astrophysics, Institute of High Energy Physics, Chinese Academy of Sciences, Beijing 100049, China}
\affiliation{
 School of Physical Sciences, University of Chinese Academy of Sciences, Beijing 100049, China 
}%
\author{Xiao-Jun Bi}
\email{bixj@ihep.ac.cn}
\affiliation{%
 Key Laboratory of Particle Astrophysics, Institute of High Energy Physics, Chinese Academy of Sciences, Beijing 100049, China}
\affiliation{
 School of Physical Sciences, University of Chinese Academy of Sciences, Beijing 100049, China 
}%
\author{Peng-Fei Yin}
\email{yinpf@ihep.ac.cn}
\affiliation{%
 Key Laboratory of Particle Astrophysics, Institute of High Energy Physics, Chinese Academy of Sciences, Beijing 100049, China}

\begin{abstract}
Many quantum gravity theories predict deviations from Lorentz invariance, a foundational principle of modern physics, at energy scales exceeding the Planck scale. The presence of Lorentz invariance violation (LIV) would modify the dispersion relations of particles in vacuum and render certain decay channels kinematically permissible. In this study, we derive stringent constraints on the energy scale of LIV in the neutrino sector by investigating the impact of neutrino decay channels on the neutrino spectrum in the presence of LIV. Utilizing the spectra derived from the joint fit of KM3NeT, IceCube, and Auger data as a benchmark, we demonstrate that to be compatible with the highest-energy neutrino event ever detected, known as event KM3-230213A, the second-order LIV energy scale must satisfy $\Lambda_2>5.0\times 10^{19}$ GeV at 90 \% confidence level. Furthermore, our analyses indicate that the current data do not impose a stringent constraint on the energy scale associated with the first-order LIV.

\end{abstract}


\maketitle
\section{Introduction}
Lorentz invariance, a fundamental symmetry within Einstein's theory of relativity, has withstood rigorous experimental tests for over a century~\cite{Kostelecky:2008ts}. However, many quantum gravity theories, which attempt to reconcile quantum mechanics with general relativity, predict deviations from Lorentz invariance at energies above the Planck scale \( E_{\mathrm{Pl}} = \sqrt{\hbar c^5/G} \approx 1.22 \times 10^{19}\,\mathrm{GeV} \)~\cite{Kostelecky:1988zi, Amelino-Camelia:1996bln, Ellis:1999rz, Amelino-Camelia:2002cqb, Magueijo:2001cr, Alfaro:2001rb, Li:2009tt, Amelino-Camelia:2008aez, Tasson:2014dfa}. Although the magnitude of Lorentz invariance violation (LIV) is expected to be extremely small at energies well below \( E_{\mathrm{Pl}} \), its impact can be amplified through cumulative effects over cosmological propagation distances. Consequently, astrophysical observations of high energy particles emitted from distant celestial sources provide uniquely powerful opportunities for probing LIV \cite{Mattingly:2005re,Wei:2021vvn,He:2022gyk,Addazi:2021xuf, Yang:2023kjq,AlvesBatista:2023wqm}.

The presence of LIV would lead to modifications in the dispersion relation of particles in vacuum, which can be expressed in natural units ($c=1$) as \cite{Borriello:2013ala}
\begin{equation}
E^2=m^2+p^2\left(1+s_n\left(\frac{E}{\Lambda_n}\right)^n\right)\;,
\end{equation}
where \( s_n = \pm 1 \) determines the sign of LIV, distinguishing between subluminal (\( s_n = -1 \)) and superluminal (\( s = +1 \)) scenarios, and \( \Lambda_n \) represents the energy scale of LIV. In the regime where particle energies are significantly lower than \( \Lambda_n \), the LIV correction is suppressed by the $n$-th power. As a result, only the first two leading orders corresponding to \( n = 1 \) and \( n = 2 \) are typically considered for LIV tests, commonly referred to as first-order and second-order LIV corrections, respectively.

Moreover, the modification of the free particle's dispersion relation in the superluminal LIV scenario facilitates the emergence of new particle decay channels that would otherwise be kinematically forbidden within the framework of standard Lorentz-invariant physics. A particular manifestation of this phenomenon is expected in superluminal neutrinos, which could potentially undergo previously prohibited decay processes, including neutrino splitting \(\nu \rightarrow \nu\nu\bar{\nu}\)~\cite{Stecker:2014oxa, Jentschura:2020nfe, Carmona:2022dtp} and vacuum electron-positron pair production \(\nu \rightarrow \nu e^+ e^-\)~\cite{Stecker:2014oxa, Cohen:2011hx, Huo:2011ve}. The unique advantage of neutrinos in probing LIV effects stems from their exceptionally weak interactions with other standard model particles, allowing high-redshift neutrinos to traverse cosmological distances to Earth with minimal attenuation or distortion. This remarkable propagation property facilitates the accumulation of LIV effects over cosmological scales, making ultrahigh-energy (UHE) neutrino observations a particularly sensitive probe for constraining the energy scale of LIV.

Recently, the KM3NeT collaboration \cite{KM3Net:2016zxf} reported the detection of an UHE neutrino event, designated as KM3-230213A~\cite{KM3NeT:2025npi}. This remarkable event is estimated to have an energy of approximately $220$ PeV, making it the highest-energy neutrino detected to date. Following this detection, several recent studies \cite{Satunin:2025uui,KM3NeT:2025mfl} have placed stringent constraints on the energy scale of LIV within the neutrino sector. These constraints are based on the consideration that a neutrino with such extreme energy would decay before reaching Earth if the LIV energy scale is insufficiently high. However, it is important to note that the constraints derived in these studies rely solely on the assumption that the mean decay length of this neutrino exceeds the distance from its source to Earth. The probability distribution concerning the distance from the neutrino source, as well as the randomness of the actual decay length of this neutrino, have not been addressed in these analyses.

In this study, we establish constraints on the energy scale of LIV, incorporating various statistical considerations into our analysis. We assume that the KM3-230213A event is part of a diffuse astrophysical neutrino spectrum. Firstly, we generate a sample of neutrinos whose energies are generated according to some benchmark neutrino spectra, derived from a joint analysis \cite{adriani2025ultra} of data collected by KM3NeT \cite{KM3NeT:2025npi,adriani2025ultra}, IceCube \cite{IceCube:2018fhm,IceCube:2020wum,IceCube:2024fxo,Abbasi:2021qfz}, and Auger \cite{PierreAuger:2023pjg}. The redshifts of these neutrinos are sampled according to a physically motivated distribution of neutrino sources \cite{Stecker:2014xja}, which aligns with the redshift distribution of the star formation rate \cite{Behroozi_2013}. Subsequently, we simulate the propagation of each neutrino toward Earth under a specific LIV energy scale, taking into account the effects of energy loss due to redshift, neutrino splitting, and vacuum electron-positron pair production. This approach allows us to determine the expected neutrino spectrum on Earth based on the energies after propagation. We then compare the expected neutrino spectrum with the full two-dimensional posterior distribution of the UHE neutrino spectrum inferred from a joint analysis \cite{adriani2025ultra} of the KM3-230213A event observed by KM3NeT, along with the nonobservations from IceCube and Auger. The presence of LIV would result in a significant suppression of the neutrino spectrum at high energies. Our analyses indicate that in order to be consistent with the UHE neutrino data, \footnote{In this paper, the terms “UHE neutrino data" or “UHE data" specifically refer to the observational results of KM3NeT, IceCube, and Auger within the energy interval $[72\text{ PeV}, 2.6\text{ EeV}]$.} the energy scale of second-order LIV in the superluminal scenario must satisfy the constraint $\Lambda_2>5.0\times 10^{19}$ GeV at 90 \% confidence level (CL). However, the current data do not impose a stringent constraint on the energy scale of first-order LIV. 

The paper is organized as follows. In Sec. \ref{Sec2}, we present the observational data and the benchmark neutrino spectra utilized in our study. In Sec. \ref{Sec3}, we describe the methods employed. The resulting constraints and corresponding discussions are presented in Sec.~\ref{Sec4}. Finally, we conclude with a summary of our findings in Sec. \ref{Sec5}.

\section{Observational data and benchmark neutrino spectrum \label{Sec2}}

\begin{figure*}
  \includegraphics[width=0.32\textwidth]{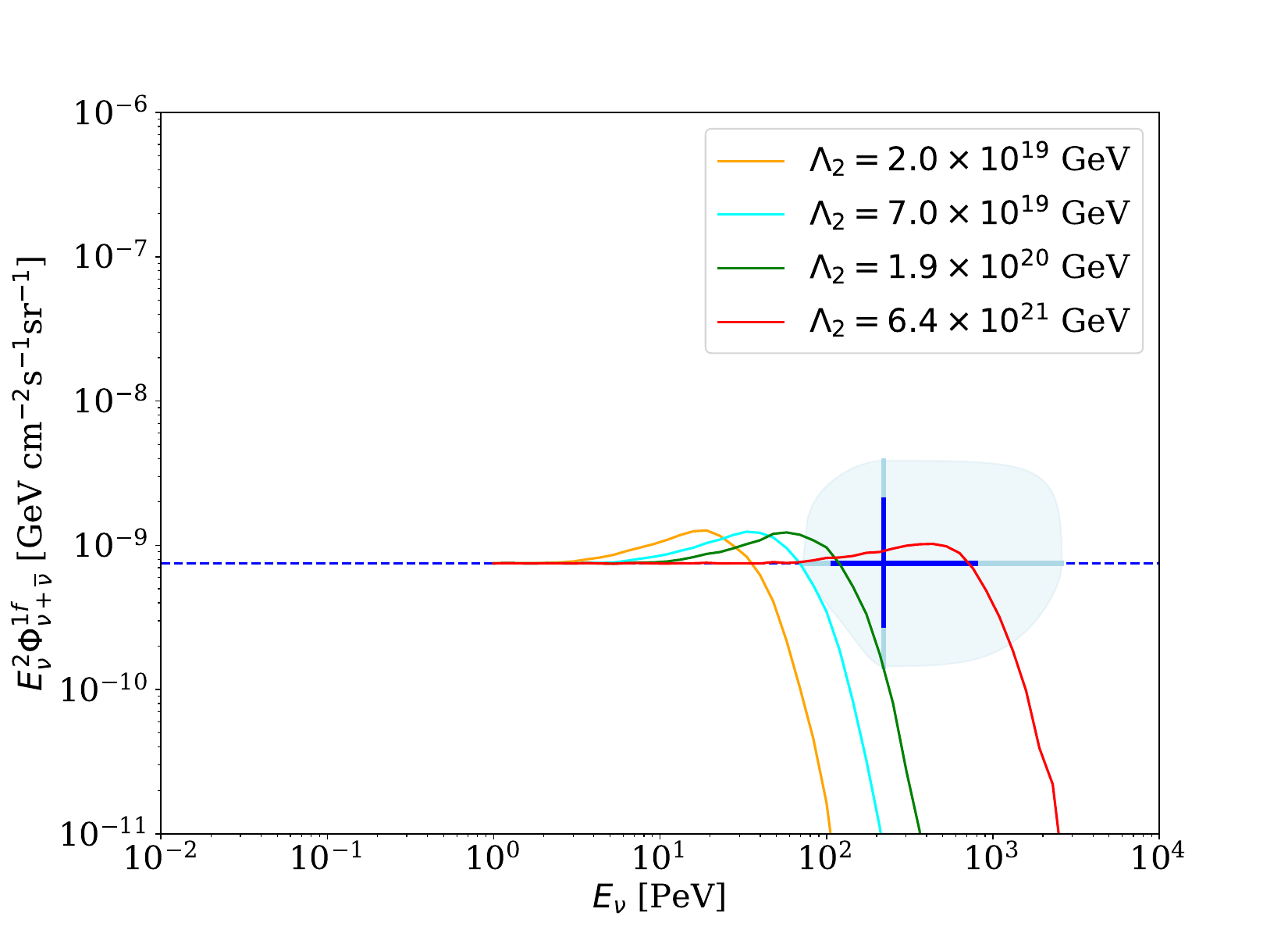}
  \includegraphics[width=0.32\textwidth]{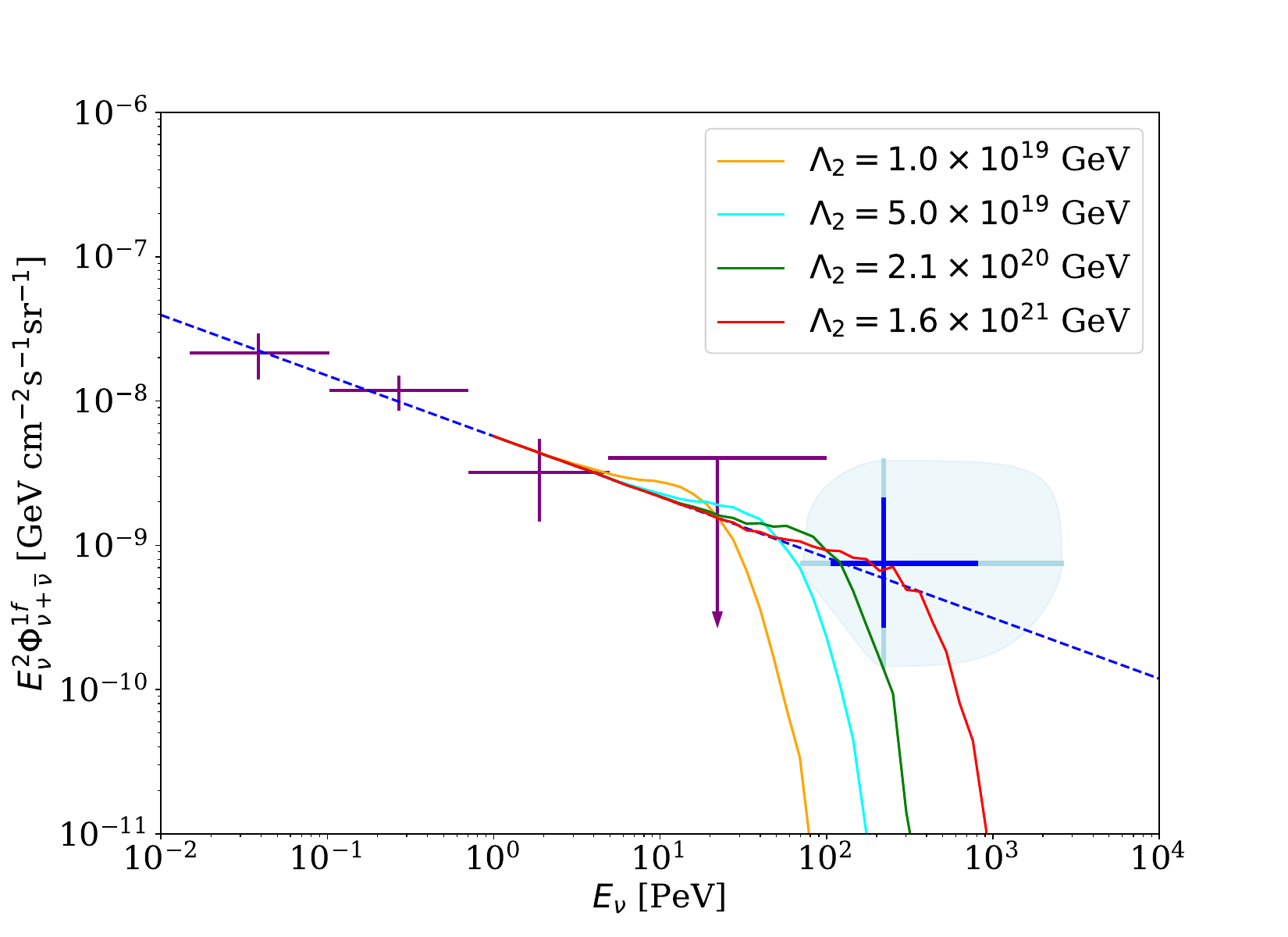}
  \includegraphics[width=0.32\textwidth]{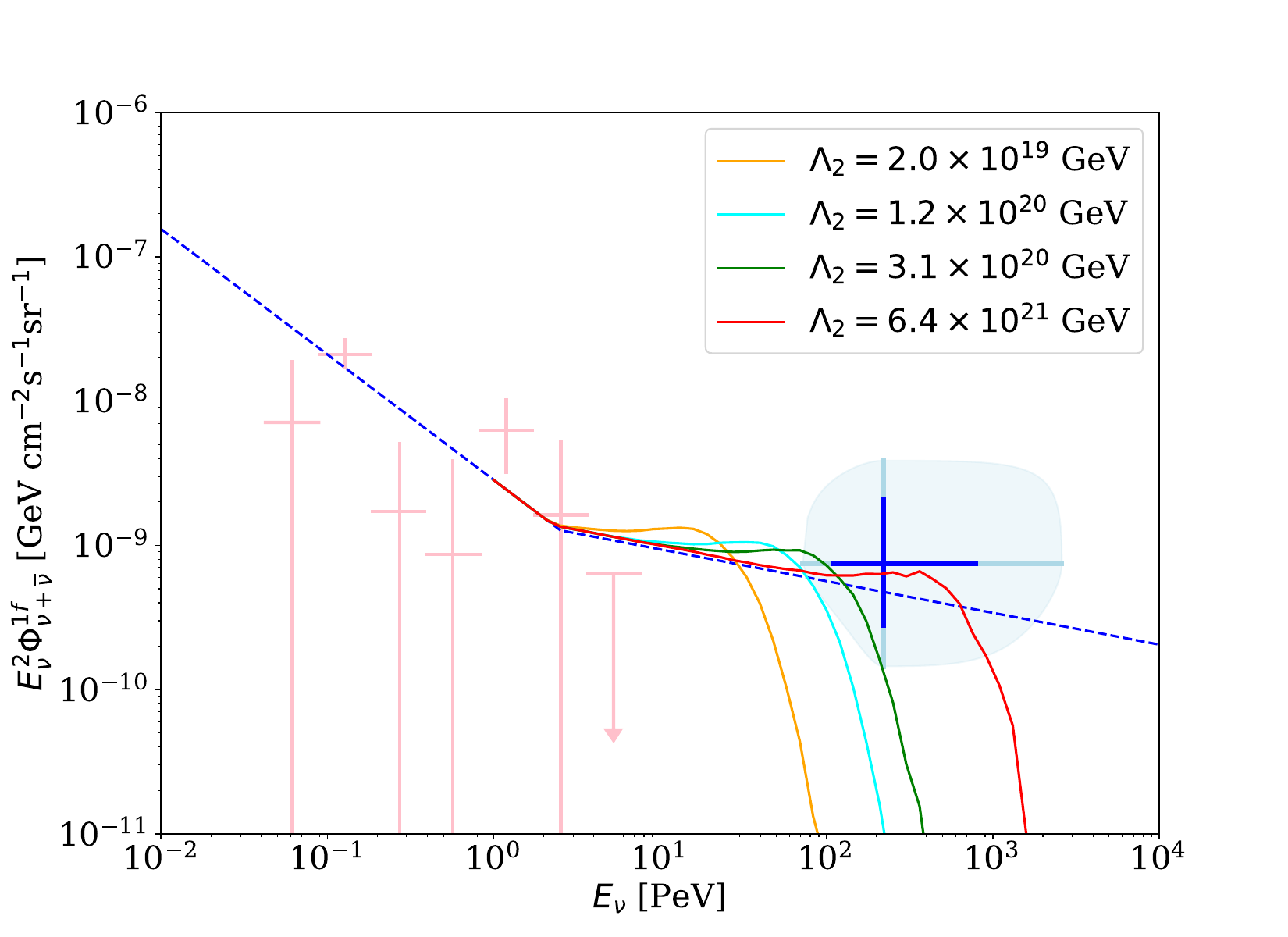}
  \caption{Expected neutrino spectra with the second-order superluminal LIV effect. The blue (light blue) crosses in all three panels denote the 68 \% (90 \%) CL intervals for both the energy of KM3-230213A event and the spectrum obtained from the joint fit of the UHE data of KM3NeT, IceCube, and Auger. The light blue shaded region represents the 90 \% confidence region for the UHE neutrino data on the $E_\nu-E_\nu^2\Phi^{1\text{f}}_{\nu+\overline{\nu}}$ plane. From left to right, the three blue dashed lines in the three panels represent the $E^{-2}$, SPL-NST, and BPL-HESE benchmark neutrino spectra in the absence of LIV, respectively. The purple (pink) crosses in the middle (right) panel represent data points for NST \cite{Abbasi:2021qfz} and HESE \cite{IceCube:2020wum}, respectively. In each panel, the different colored solid lines depict the expected neutrino spectra under various second-order LIV energy scales in the superluminal scenario.}
  \label{figure_1}
\end{figure*}

The best estimated energy of the UHE neutrino event KM3-230213A, detected by KM3NeT, is 220 PeV, with a 68 \% (90 \%) CL interval of [110 PeV, 790 PeV] ([72 PeV, 2.6 EeV]). Assuming a $E^{-2}$ spectrum within the interval of [72 PeV, 2.6 EeV], Ref. \cite{adriani2025ultra} conducted a joint analysis of this single event observed by KM3NeT \cite{KM3NeT:2025npi}, along with nonobservations from IceCube \cite{IceCube:2018fhm} and Auger \cite{PierreAuger:2023pjg} within this energy interval. From this analysis, they derived a per-flavor neutrino spectrum of $E_\nu^2\Phi^{1\text{f}}_{\nu+\overline{\nu}}(E_\nu)=7.5^{+13.1}_{-4.7}\times 10^{-10}$ GeV cm$^{-2}$ s$^{-1}$ sr$^{-1}$ \cite{adriani2025ultra}. By assuming a Gaussian distribution for the logarithm of this neutrino spectrum, we derive the 90 \% CL interval for it as $[1.45,38.66]\times 10^{-10}$ GeV cm$^{-2}$ s$^{-1}$ sr$^{-1}$. The 68 \% (90 \%) CL intervals for both the energy of KM3-230213A event and the spectrum obtained from the joint fit are illustrated in Fig. \ref{figure_1} with blue (light blue) cross. We consider the area enclosed by four quarter ellipses centered at $(220\text{ PeV}, 7.5\times 10^{-10}\text{ GeV cm}^{-2}\text{ s}^{-1}\text{ sr}^{-1})$,  with the endpoints of the light blue cross serving as vertices, on the $E_\nu-E_\nu^2\Phi^{1\text{f}}_{\nu+\overline{\nu}}$ plane to represent the 90 \% confidence region for the UHE neutrino data. This region is depicted as the light blue shaded area in Fig. \ref{figure_1} on a logarithmic coordinate plot. If the curve of a neutrino spectrum intersects this region, we consider it to be consistent with the UHE data at a CL of at least 90 \% \cite{2018ReprintOM, DEMAESSCHALCK20001}. A detailed description of this method is provided in Appendix~\ref{App_1}. For comparison, we also conduct an analysis that disregards the energy uncertainty of the data, simply requiring that the expected neutrino spectrum does not fall below the lower bound of the 90\% CL interval at 220 PeV.

We refer to the expected neutrino spectrum in the absence of LIV as the benchmark spectrum. In this study, we examine three distinct forms of this spectrum. The first form is the $E^{-2}$ spectrum, with its normalization derived from the joint fitting of KM3NeT, IceCube, and Auger UHE data, as mentioned earlier. Specifically, we adopt the best-fit result reported in Ref. \cite{adriani2025ultra}
\begin{equation}
   \Phi^{1\text{f}}_{\nu+\overline{\nu}}(E_\nu)=7.5\times 10^{-10}\text{ GeV cm}^{-2}\text{ s}^{-1}\text{ sr}^{-1}E^{-2}_\nu.
   \label{E_2}
\end{equation}
This spectrum is depicted as the blue dashed line in the left panel of Fig. \ref{figure_1}.

The second form of the benchmark spectrum we consider is a single power law (SPL), expressed as
\begin{equation}
    \Phi^{1\text{f}}_{\nu+\overline{\nu}}(E_\nu)=\phi\times \left( \frac{E_\nu}{100\text{ TeV}}\right)^{-\gamma_1},
    \label{SPL}
\end{equation}
where the normalization $\phi$ and spectral index $\gamma_1$ are obtained in Ref. \cite{adriani2025ultra} through a fit of the UHE neutrino data from KM3NeT, IceCube, and Auger, along with the high-energy (below tens of PeV) measurements from IceCube. The IceCube high-energy measurements include three samples: High-Energy Starting Events (HESE) \cite{IceCube:2020wum}, Enhanced Starting Track Event Selection (ESTES) \cite{IceCube:2024fxo}, and Northern-Sky Tracks (NST) \cite{Abbasi:2021qfz}. Each of these samples is independently combined with the UHE data to obtain three sets of best-fit values for $\phi$ and $\gamma_1$, as reported in Ref. \cite{adriani2025ultra}. Accordingly, we refer to these three SPL benchmark spectra as SPL-HESE, SPL-ESTES, and SPL-NST, respectively. The SPL-NST spectrum is depicted as the blue dashed line in the middle panel of Fig. \ref{figure_1}, with the purple crosses representing the NST data points.

The third form of the benchmark spectrum is a broken power law (BPL), expressed as
\begin{equation}
    \Phi^{1\text{f}}_{\nu+\overline{\nu}}(E_\nu)=\phi\times\left\{\begin{aligned}
        &\left(\frac{E_\nu}{100\text{ TeV}}\right)^{-\gamma_1},\, E_\nu\leq E_b,\\
        &\left(\frac{E_\nu}{E_b}\right)^{-\gamma_2}\left(\frac{E_b}{100\text{ TeV}}\right)^{-\gamma_1},\,E_\nu>E_b.\end{aligned}\right.
    \label{BPL}
\end{equation}
In this case, there are three sets of best-fit values for $\phi$, $\gamma_1$, $\gamma_2$, and $E_b$ in Ref. \cite{adriani2025ultra}, also obtained by separately fitting the HESE, ESTES, and NST samples with the UHE data. Similarly, we refer to these three BPL benchmark spectra as BPL-HESE, BPL-ESTES, and BPL-NST, respectively. The BPL-HESE spectrum is depicted as the blue dashed line in the right panel of Fig. \ref{figure_1}, with the pink crosses representing the HESE data points. All detailed fitting results, including those for the $E^{-2}$, SPL, and BPL spectra, can be found in Ref. \cite{adriani2025ultra}.

\section{Method \label{Sec3}}
In this work, we establish constraints on the superluminal LIV energy scale by investigating the suppressive effect of LIV on the high-energy end of the expected neutrino spectrum on Earth \cite{Stecker:2014xja, Stecker:2014oxa}. In the absence of LIV, the expected neutrino spectrum on Earth is the same as one of the above-mentioned benchmark spectra. With the exception of the SPL-HESE spectrum, which we exclude from our subsequent analyses, all other benchmark spectra intersect the 90 \% confidence region of the UHE data, as illustrated in Fig. \ref{figure_1}. However, under a specific superluminal LIV energy scale, the introduction of decay channels results in significant energy loss for neutrinos, particularly at the high-energy end. This leads to a pronounced decrease in the expected neutrino spectrum on Earth compared to the scenario without LIV. When this expected neutrino spectrum tangentially exits the 90 \% confidence region of the UHE data, the corresponding LIV energy scale can be considered as a lower limit on the LIV energy scale at 90 \% CL. 
 
In our analysis, we utilize a Monte Carlo method to simulate the energy loss of neutrinos during their propagation. To achieve this, we construct an event sample comprising $10^7$ neutrinos. Since the initial spectrum of the emitted neutrinos is unknown, we assume that in the absence of LIV, the neutrino spectrum observed on Earth follows one of the aforementioned benchmark spectral forms. Specifically, we sample $10^7$ energies $E^{(i)}\,(i=1,2,\cdots, 10^7)$ within the range of $[1,10^4]$ PeV according to the benchmark spectrum. The interpretation of these energies depends on the specific form of the benchmark spectrum, which will be elaborated upon in Sec. \ref{Redshift}. Additionally, the redshifts $z^{(i)}_\text{emit}\,(i=1,2,\cdots, 10^7)$ of these neutrinos, as they are emitted from their sources, are sampled within the range of $[0,7]$ according to the redshift distribution described in Ref. \cite{Stecker:2014xja}, which aligns with the redshift distribution of the star formation rate \cite{Behroozi_2013}.

\subsection{Redshift effect}
\label{Redshift}

In the absence of LIV, neutrinos experience energy loss due to cosmological redshift during their propagation. The energy loss can be expressed as 
\begin{equation}
    \frac{\partial \ln E_\nu}{\partial t}=-H(z)=-H_0\sqrt{\Omega_m (1+z)^3+\Omega_\Lambda},
    \label{redshift}
\end{equation}
where the cosmological constants are taken to be $H_0=67.8$ km s$^{-1}$ Mpc$^{-1}$, $\Omega_\Lambda=0.7$, and $\Omega_m=1-\Omega_\Lambda=0.3$, respectively. 

In the absence of LIV, when a neutrino is emitted from its source with energy $E_\text{emit}$ and $z_\text{emit}$, its energy upon reaching Earth is reduced to $E_\text{redshift}=E_\text{emit}/(1+z_\text{emit})$ due to the redshift effect. Consequently, if we assume that the emitted energies of the neutrinos in our event sample adhere to the benchmark spectrum, the expected neutrino spectrum on Earth may deviate from the benchmark spectrum due to the redshift effect, even in the absence of LIV. However, for an initial sample generated with the power-law benchmark spectrum, it can be shown that the expected spectrum on Earth still follows a power law with the same spectral index when LIV is absent. We have confirmed this conclusion through actual Monte Carlo simulations, as described in Appendix \ref{App_2}. Therefore, for the $E^{-2}$ and SPL benchmark spectra, we can identify the emitted energy $E^{(i)}_\text{emit}$ to be the previously sampled energy $E^{(i)}$. This allows us to establish the initial condition for our event sample as $\left\{E^{(i)}_\text{emit}=E^{(i)},z^{(i)}_\text{emit}\,|\,i=1,2,\cdots,10^7\right\}$ in the $E^{-2}$ and SPL benchmark spectrum cases.

Conversely, if we assume that the emitted energies of the neutrinos in our event sample conform to a BPL benchmark spectrum, the expected spectrum on Earth would deviate from this BPL spectrum. This can be understood by recognizing that the redshifted break energy $E_{b,\text{expected}}=E_b/(1+z_\text{emit})$ varies with redshift. Consequently, for the BPL benchmark spectrum case, to ensure that the expected neutrino spectrum on Earth aligns with the benchmark spectrum, we multiply the previously sampled energy $E^{(i)}$ of each neutrino by the corresponding redshift factor $1+z^{(i)}_\text{emit}$ to obtain its emitted energy $E^{(i)}_\text{emit}=(1+z^{(i)}_\text{emit})E^{(i)}$. This approach allows us to establish the initial condition for our event sample as $\left\{E^{(i)}_\text{emit}=(1+z^{(i)}_\text{emit})E^{(i)},z^{(i)}_\text{emit}\,|\,i=1,2,\cdots,10^7\right\}$ in the context of the BPL benchmark spectrum. We have verified through Monte Carlo simulations that, in the absence of LIV, the expected neutrino spectrum on Earth indeed conforms to the same BPL spectrum.

\subsection{Superluminal neutrino decay}
In the presence of LIV, superluminal neutrinos undergo decay in vacuum primarily through two processes: vacuum electron-positron pair production $\nu\to\nu e^+e^-$ and neutrino splitting $\nu\to \nu\nu\bar{\nu}$. The threshold for the $\nu\to\nu e^+e^-$ process is given by \cite{Borriello:2013ala,Stecker:2014xja,Stecker:2014oxa}
\begin{equation}
    E_\text{th}(E_\nu)=2m_e\left(\frac{E_\nu}{\Lambda_n}\right)^{-n/2},
\end{equation}
where $m_e$ is the electron mass and $E_\nu$ is the neutrino energy.
The threshold for $\nu\to \nu\nu\bar{\nu}$ is significantly lower than the energy range under consideration \cite{Stecker:2014oxa}, allowing us to neglect it. Ignoring the contribution of charged current $W$-exchange channels \cite{Stecker:2014xja}, the decay rate of the $\nu\to\nu e^+e^-$ process \cite{cohen2011pair,Stecker:2014xja} is calculated by
\begin{equation}
    \Gamma_{\nu\to\nu e^+e^-}(E_\nu)=\frac{1}{14}\frac{G^2_FE^5_\nu}{192\pi^3}\left(\frac{E_\nu}{\Lambda_n}\right)^{3n},
\end{equation}
where $G_F$ is the Fermi coupling constant. The decay rate of the $\nu\to \nu\nu\bar{\nu}$ process is three times the above result \cite{Stecker:2014oxa}
\begin{equation}
    \Gamma_{\nu\to\nu\nu\bar{\nu}}(E_\nu)=\frac{3}{14}\frac{G^2_FE^5_\nu}{192\pi^3}\left(\frac{E_\nu}{\Lambda_n}\right)^{3n}.
\end{equation}

 When neutrinos decay during their propagation, they generate lower-energy neutrinos. In the $\nu\to\nu e^+e^-$ process, we assume that the neutrino loses 78 \% of its energy \cite{cohen2011pair}. In the $\nu\to\nu\nu\bar{\nu}$ process, the original neutrino transforms into three new neutrinos \footnote{In the following discussions, unless otherwise specified, we will collectively refer to neutrinos and antineutrinos as “neutrinos," since the detectors cannot distinguish between the two.}, with each of the resulting neutrinos carrying approximately $1/3$ of the original energy.

 \subsection{Simulation of neutrino propagation process}
 For a given LIV energy scale $\Lambda_n$, we simulate the energy loss due to both the redshift and decay effects for the neutrinos events $\left\{E_\text{emit}^{(i)},z^{(i)}_\text{emit}\,|\,i=1,2,\cdots,10^7\right\}$ during propagation. The distance of a neutrino at redshift $z$ must travel before reaching Earth  is expressed as (with $c=1$)
 \begin{equation}
     D(z)=\frac{1}{H_0}\int_0^z\frac{dz^\prime}{(1+z^\prime)\sqrt{\Omega_m(1+z^\prime)^3+\Omega_\Lambda}}.
     \label{D_z}
 \end{equation}
Due to the process of neutrino splitting, the total number of neutrinos during the propagation process generally exceeds $10^7$. We denote the state of the $j$-th neutrino at any moment during its propagation by $E^{(j)},z^{(j)}$, and $D^{(j)}\,(j=1,2,\cdots)$, where $D^{(j)}$ and $z^{(j)}$ can be mutually determined through Eq. \ref{D_z}. 

The propagation step of the neutrinos is set to $\Delta D=1$ Mpc, and we have verified that the results remain consistent with smaller propagation steps. In each propagation step for the neutrino labeled $j$, we sequentially update its distance, energy, and redshift as follows:
\begin{equation}
    \begin{aligned}
        &D^{(j)}\to D^{(j)}-\Delta D,\\
        &E^{(j)}\to E^{(j)}\exp\left[-\Delta D H_0\sqrt{\Omega_m(1+z^{(j)})^3+\Omega_\Lambda}\right],\\
        &z^{(j)}\to z\left(D^{(j)}-\Delta D\right).\\
    \end{aligned}
    \label{step_1}
\end{equation}
Here, the energy update, which is derived from Eq. \ref{redshift},  accounts for the energy lost by the neutrino due to redshift during this step of propagation. The function $z(D)$ serves as the inverse of Eq. \ref{D_z}, linking a specific distance $D$ to its corresponding redshift value. This inverse function can be expressed analytically as
\begin{equation}
    z(D)=\frac{(4\Omega_\Lambda)^{1/3}e^{\sqrt{\Omega_\Lambda}H_0D}}{\left[1+\sqrt{\Omega_\Lambda}-(1-\sqrt{\Omega_\Lambda})e^{3\sqrt{\Omega_\Lambda}H_0D}\right]^{2/3}}-1.
\end{equation}
Accordingly, the term $z\left(D^{(j)}-\Delta D\right)$ represents the updated redshift value associated with the updated distance.

The effect of neutrino decay should be considered during the same propagation step. However, this effect varies depending on whether it is first-order ($n=1$) or second-order ($n=2$) LIV. In the case of second-order LIV, $CPT$ symmetry is preserved \cite{Stecker:2014oxa}, resulting in the same signs of LIV for both neutrinos and antineutrinos. Assuming that both are superluminal, both of them are likely to decay in the propagation step described by Eq. \ref{step_1}. The probabilities for the occurrences of the processes $\nu\to\nu e^+e^-$ and $\nu\to\nu\nu\bar{\nu}$ in this step are given by
\begin{equation}
\begin{aligned}
    P(\nu\to\nu e^+e^-)=&\theta\left[E^{(j)}-E_\text{th}(E^{(j)})\right]\\
    &\times\left[1-e^{-\Gamma_{\nu\to\nu e^+e^-}(E^{(j)})\Delta D}\right],
\end{aligned}
\end{equation}
and
\begin{equation}
    P(\nu\to\nu\nu\bar{\nu})=1-e^{-\Gamma_{\nu\to\nu\nu\bar{\nu}}(E^{(j)})\Delta D},
\end{equation}
respectively. Here, the $\theta(x)$ is the Heaviside step function, which accounts for the threshold effect of the process $\nu\to\nu e^+e^-$. If the process $\nu\to\nu e^+e^-$ occurs, we reduce the energy of this neutrino to $0.22 E^{(j)}$. If the process $\nu\to\nu\nu\bar{\nu}$ occurs, we replace the original neutrino with three neutrinos, each carrying an energy of $E^{(j)}/3$. 

In the context of first-order LIV, the $CPT$ symmetry is also violated \cite{Stecker:2014oxa}. This implies that if the neutrino is superluminal, the antineutrino must be subluminal, and vice versa. In this scenario, there is always a fraction of particles that remain subluminal and do not undergo decay, with energy loss occurring solely due to redshift. Consequently, the initial ratio of superluminal to subluminal particles plays a significant role in influencing the results. In our analysis, this ratio is assumed to be 1:1, and its specific value does not affect our main conclusion. Furthermore, during the process of neutrino splitting, a superluminal particle generates two superluminal particles and one subluminal particle. The resulting subluminal particle does not undergo decay during its subsequent propagation. This characteristic has been taken into account in our simulation.

After all the neutrinos have propagated to Earth, we conduct a statistical analysis on the obtained energies to determine the expected neutrino spectrum on Earth. Given that low-energy neutrinos remain unaffected by  LIV effects, we normalize the expected spectrum at $E_\nu=1$ PeV to the value of the corresponding benchmark spectrum at the same energy. As shown in Fig. \ref{figure_1}, the spectral shape around 1 PeV indeed remains unchanged after propagation, confirming that our choice of normalization point is reasonable.

\section{Results and discussions \label{Sec4}}
\begin{figure}
\includegraphics[width=\linewidth]{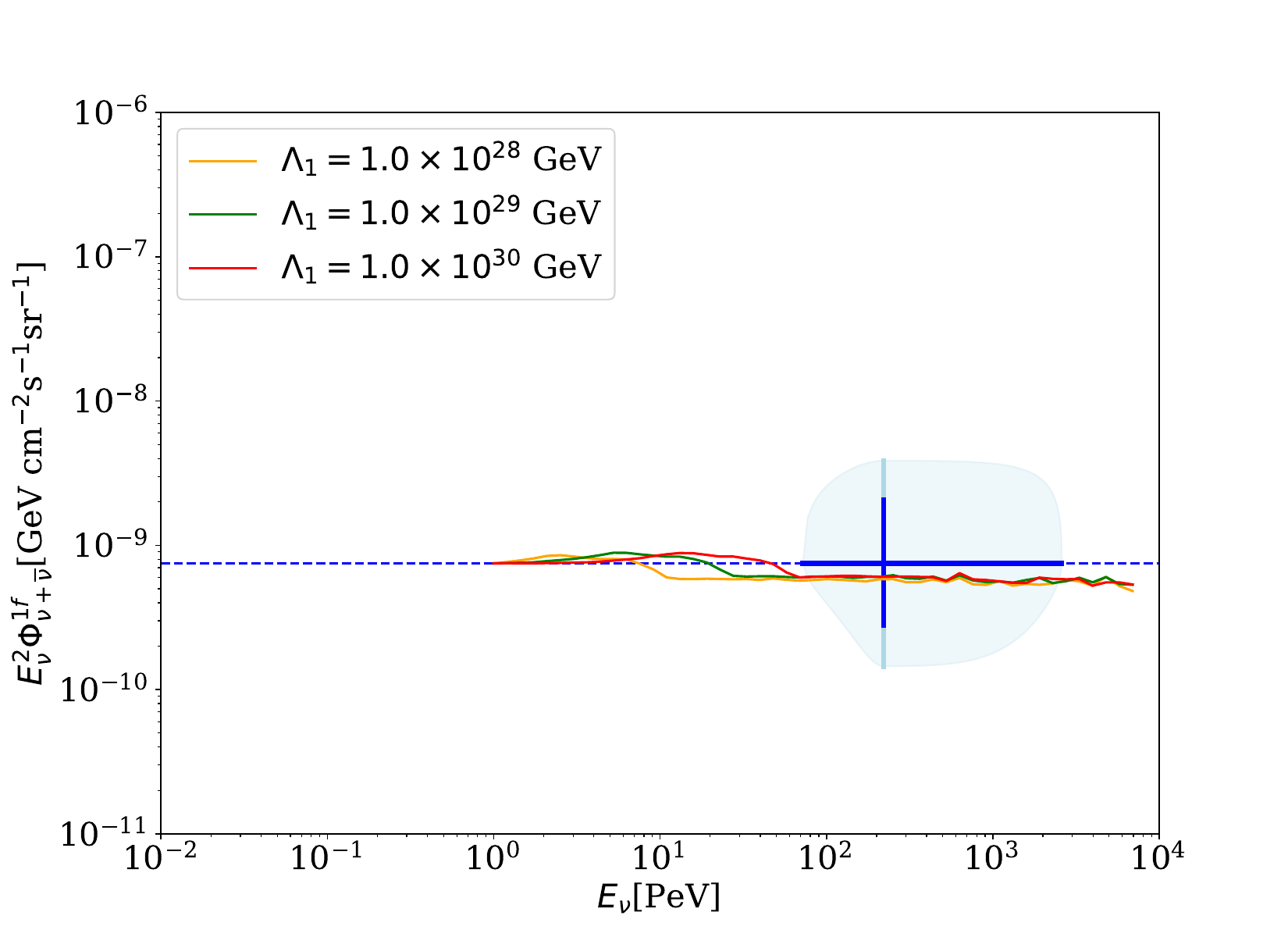}
  \caption{Impact of first-order LIV on the expected neutrino spectrum on Earth, for a benchmark $E^{-2}$ spectrum. In our event sample, the initial ratio of superluminal to subluminal particles is set to be $1:1$. The different colored solid lines represent the results for various LIV energy scales. The meanings of the blue and light blue crosses, along with the light blue shaded region, are the same as those depicted in Fig. \ref{figure_1}.}
  \label{figure_2}
\end{figure}

In the case of second-order LIV, assuming that both neutrino and antineutrino are superluminal, the expected neutrino spectra on Earth are illustrated in Fig. \ref{figure_1} as solid lines of various colors, each corresponding to different LIV energy scales. The three panels of this figure present results associated with the $E^{-2}$, SPL-NST, and BPL-HESE benchmark spectra, respectively. A notable feature of the expected spectrum is the presence of a pileup prior to the steep dropoff, which emerges naturally from the decay of higher-energy neutrinos into lower-energy neutrinos. The contribution to the pileup from the process $\nu\to\nu\nu\bar{\nu}$ is greater than that from the process $\nu\to\nu e^+e^-$. This is because the former process generates three new neutrinos with lower energies, whereas the latter leads to only one new neutrino with a lower energy. This phenomenon has also been discussed in Ref. \cite{Stecker:2014oxa}.
\begin{table*}[htbp]
    \centering
    \caption{90\% CL lower constraints on the second-order LIV energy scale derived in this study. The first row specifies the form of neutrino benchmark spectrum used, while the second row  identifies the IceCube data samples employed in fitting the benchmark spectrum. The third and fourth rows present the 90\% CL lower bounds of the second-order LIV energy scale, obtained with and without considering the energy uncertainty of KM3-230213A, respectively. All results are expressed in units of $10^{19}$ GeV.}
    \begin{ruledtabular}
    \begin{tabular}{c|ccccccc}
    &$E^{-2}$&\multicolumn{3}{c}{SPL}&\multicolumn{3}{c}{BPL}\\
    &&HESE& ESTES&NST&HESE& ESTES&NST\\
    \hline
    Considering energy uncertainty&7.0&...&20.0&5.0&12.0&10.0&8.0\\
    Ignoring energy uncertainty&19.0&...&39.0&21.0 &31.0&31.0&29.0\\
    \end{tabular}
    \end{ruledtabular}
    \label{Tab1}
\end{table*}

In the case of a $E^{-2}$ benchmark spectrum, when the LIV energy scale is $7.0\times 10^{19}$ GeV, the resulting neutrino spectrum becomes tangent to the 90 \% confidence region of the UHE neutrino data, as shown by the cyan solid line in the left panel of Fig. \ref{figure_1}. This enables us to establish a lower limit for the second-order LIV energy scale at $7.0\times 10^{19}$ GeV at the 90 \% CL. Similarly, for the SPL-NST, as represented by the cyan solid line in the middle panel of Fig. \ref{figure_1}, and for the SPL-ESTES cases, the corresponding 90 \% CL lower limits are determined to be $5.0\times 10^{19}$ GeV and $2.0\times 10^{20}$ GeV, respectively. Additionally, as the SPL-HESE benchmark spectrum does not intersect with the 90 \% confidence region, we choose not to utilize it to impose constraints. The 90 \% CL lower limits obtained for the BPL-HESE, corresponding to the cyan solid line in the right panel of Fig. \ref{figure_1}, as well as for the BPL-NST and BPL-ESTES benchmark spectra, are $1.2\times 10^{20}$ GeV, $8.0\times 10^{19}$ GeV, and $1.0\times 10^{20}$ GeV, respectively. Based on these results, we can conservatively establish a 90 \% CL lower limit on the second-order LIV energy scale as $5.0\times 10^{19}$ GeV. All these constraints are summarized in the third row of Table~\ref{Tab1}.

Using the $E^{-2}$ benchmark spectrum as a case study, the simulation results for the first-order LIV scenario are presented in Fig. \ref{figure_2}. The lines of various colors represent the expected neutrino spectra on Earth, calculated for different LIV energy scales. It is evident that, owing to the significant presence of subluminal particles, these spectra do not exhibit a substantial decrease at the high-energy end, consistently passing through the 90 \% confidence region of the UHE data. This indicates that the current observational data do not impose constraints on the first-order LIV energy scale.

To better understand the impact of the energy uncertainty associated with the KM3-230213A event on the constraint results, we perform an alternative analysis under the assumption that the event's energy is exactly 220~PeV, disregarding its uncertainty. In this approach, the 90\% CL lower limit on the second-order LIV energy scale is directly taken to be the energy scale at which the expected spectrum on Earth reaches the lower edge of the 90\% CL interval of the observed data point at 220 PeV, as indicated by the green solid lines in Fig.~\ref{figure_1}. The resulting constraint values are summarized in the fourth row of Table~\ref{Tab1}. A comparison of the third and fourth rows of Table~\ref{Tab1} shows that incorporating the energy uncertainty of KM3-230213A weakens the constraints by a factor of approximately $2-4$. An event with a true energy below 220~PeV would naturally yield weaker constraints on LIV compared to those derived under the assumption of a fixed energy of 220 PeV.

Many previous studies have placed constraints on LIV in the neutrino sector, enabling a meaningful comparison with our results. Since many of these studies present their limits in terms of the parameter $\delta = (v/v_0)^2 - 1 \simeq (E_\nu/\Lambda_n)^n$, we convert all previous constraints into direct limits on $\Lambda_n$ to facilitate comparison. Here, we focus  specifically on the case of second-order LIV. Using data from SN1987A, Stodolsky obtained a constraint of $\Lambda_2\gtrsim 100$ GeV \cite{STODOLSKY1988353}, considering that the typical energy of supernova neutrinos is $E_\nu\simeq 10$ MeV \cite{Borriello:2013ala}. Based on two cascade neutrino events with energies around $1$ PeV observed by IceCube during 2011–2012, Borriello $et$ $al$. derived a strong constraint of $\Lambda_2 \gtrsim 10^{15}$ GeV \cite{Borriello:2013ala}. Stecker and Scully analyzed the energy spectrum of neutrinos observed by IceCube below $\sim 2$ PeV and obtained a constraint of $\Lambda_2 \gtrsim 10^{16}$ GeV \cite{Stecker:2014xja}. In the context of the KM3-230213A event, other studies have reported constraints comparable to our results. Satunin \cite{Satunin:2025uui} and the KM3NeT Collaboration \cite{KM3NeT:2025mfl} both derived limits around $\Lambda_2 \gtrsim 10^{19}$ GeV. Therefore, the exceptionally high energy of KM3-230213A imposes significantly more stringent constraints on the second-order LIV energy scale compared to previous observations.

\section{Conclusion \label{Sec5}}
In this study, we utilize the recently observed UHE neutrino event KM3-230213A, along with the energy spectra obtained from a joint fit of data from KM3NeT, IceCube, and Auger, to impose constraints on the LIV energy scale. Our analysis indicates that the current data do not impose strict constraints on the first-order LIV energy scale. For the second-order LIV energy scale, under the assumption that both neutrino and antineutrino are superluminal, we establish a lower limit of $5.0\times 10^{19}$ GeV at a 90 \% CL.

\acknowledgments
We thank En-Sheng Chen for helpful discussions. This work is supported by the National Natural Science Foundation of China under Grants No. 12175248 and No. 12447105.

\section{Data availablity \label{Sec5}}
The data that support the findings of this article are openly available \cite{KM3NeT:2025npi, adriani2025ultra}.

%

\begin{appendix}
\section{The criterion of 90\% constraints } \label{App_1}
To illustrate the methodology, we take the second-order LIV as an example, where the sole parameter to be constrained is the LIV energy scale $\Lambda_2$. Each specific value of $\Lambda_2$ determines an expected spectrum on Earth, denoted as $E_\nu^2\Phi^{1\text{f}}_{\nu+\overline{\nu},\Lambda_2}(E_\nu)$. To quantify the compatibility between the KM3-230213A data point and the theoretical curve depending on $\Lambda_2$ in the $E_\nu - E_\nu^2\Phi^{1\text{f}}_{\nu+\overline{\nu}}$ plane, we introduce a distance as follows
\begin{equation}
\begin{aligned}
    &d(\Lambda_2)=\min_{E_\nu}\left\{\left[\frac{(E_\nu-E_\text{best})^2}{(E_\text{best}-E_\text{lower})^2}\right.\right.\\
    &\left.\left.+\frac{\left[E_\nu^2\Phi^{1\text{f}}_{\nu+\overline{\nu},\Lambda_2}(E_\nu)-E_\text{best}^2\Phi^{1\text{f}}_{\nu+\overline{\nu},\text{best}}(E_\text{best})\right]^2}{\left[E_\text{best}^2\Phi^{1\text{f}}_{\nu+\overline{\nu},\text{best}}(E_\text{best})-E_\text{best}^2\Phi^{1\text{f}}_{\nu+\overline{\nu},\text{lower}}(E_\text{best})\right]^2}\right]^{\frac{1}{2}}\right\},
\end{aligned}
\label{Eq_A1}
\end{equation}
where $E_\text{best}$ and $E_\text{lower}$ represent the best estimate and lower bound of the 90\% CL interval for the energy of the KM3-230213A event, respectively, and $\Phi^{1\text{f}}_{\nu+\overline{\nu},\text{best}}(E_\text{best})$ and $\Phi^{1\text{f}}_{\nu+\overline{\nu},\text{lower}}(E_\text{best})$ denote the best estimate and lower bound of the 90\% CI for the observed flux at $E_\text{best}$, respectively. It is important to note that Eq. \ref{Eq_A1} is formulated for the case in which the point on the expected spectrum curve minimizing the distance always lies in the lower-left quadrant relative to the data point—a configuration precisely encountered in our analysis, as illustrated in Fig. \ref{figure_1}.

Our definition of distance in Eq. \ref{Eq_A1} is inspired by the concept of the Mahalanobis distance \cite{2018ReprintOM,DEMAESSCHALCK20001}. The Mahalanobis distance can be formally defined by setting the denominators in Eq. \ref{Eq_A1} with the squares of the standard deviations of the data point’s horizontal and vertical coordinates, in the case of a Gaussian distribution. Furthermore, the Mahalanobis distance follows a $\chi^2$ distribution. However, since the estimated energy of KM3-230213A does not follow a standard Gaussian distribution, the standard deviation cannot be directly inferred from the experimentally reported CL intervals. Consequently, in Eq. \ref{Eq_A1}, we directly use the 90\% CL interval of the energy to define the corresponding term in the distance.

We use $d(\Lambda_2)$ as a measure of the compatibility between the expected spectrum curve (or, equivalently, the sole parameter $\Lambda_2$) and the data point. Thus, the values of 
$\Lambda_2$ that satisfy $d(\Lambda_2)\leq 1$ are considered consistent with the data at a CL of at least 90\%. Consequently, the value of $\Lambda_2$ that satisfies $d(\Lambda_2)=1$ is defined as the lower bound of $\Lambda_2$ at the 90\% CL. Geometrically, the set of points satisfying $d=1$ in the $E_\nu-E_\nu^2\Phi^{\text{1\text{f}}}_{\nu+\overline{\nu}}$ plane forms an ellipse, which corresponds to the boundary of the light blue shaded region shown in Fig. \ref{figure_1}. In practice, we take this shaded region as the 90\% confidence region of the data point. The energy scale $\Lambda_2$ for which the expected spectrum curve is tangent to this region satisfies $d(\Lambda_2)=1$.

\section{The impact of redshift on power-law spectra} \label{App_2}
Even in the absence of LIV, emitted neutrinos lose energy during propagation due to cosmological redshift, which can potentially cause the spectrum on Earth to deviate from the originally emitted spectrum. However, in this section, we demonstrate that if the emitted spectrum follows a power-law distribution, the expected spectrum on Earth will retain the same power-law form in the absence of LIV. This conclusion is validated through both theoretical analysis and Monte Carlo simulations.

We begin with a theoretical derivation, which necessitates the introduction of specific notation. Let $\Phi(E)$ denote the overall spectrum of neutrinos emitted across all redshifts, without considering propagation effects. The spectrum of neutrinos emitted specifically at redshift $z$ is denoted as $\Phi_z$, which is assumed to be independent of redshift. For a power-law spectrum, we define the spectral index as $\alpha$, such that $\Phi_z(E)\propto E^{-\alpha}$. Let $P(z)$ represent the redshift distribution of neutrino sources. The overall spectrum can be given by $\Phi(E)\propto \int P(z)\Phi_z(E)dz\propto E^{-\alpha}\int P(z)dz=E^{-\alpha}$, which retains a power-law form.

\begin{figure}[h]
\includegraphics[width=\linewidth]{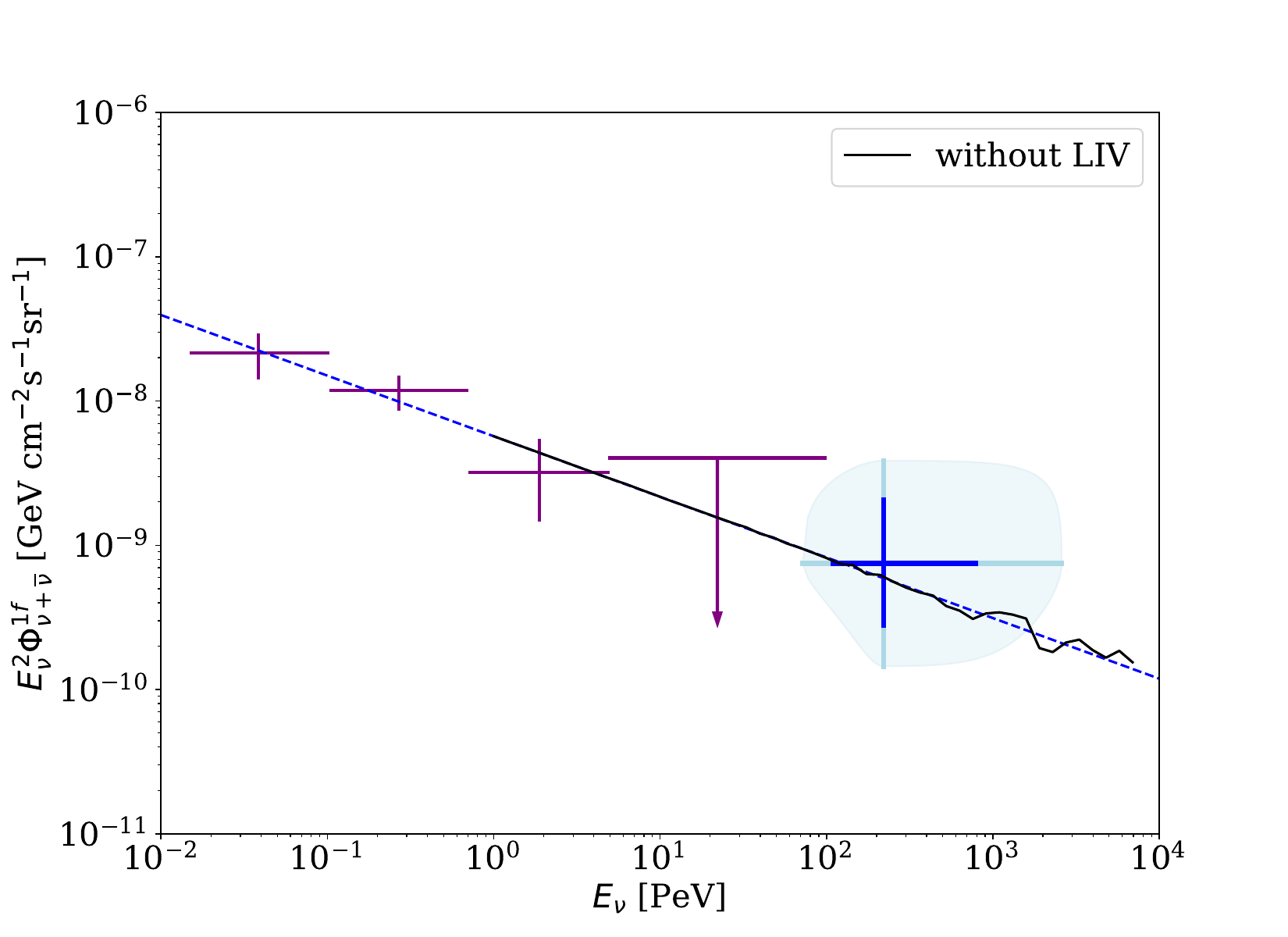}
  \caption{Impact of redshift effects on the expected neutrino spectrum on Earth, for the SPL-NST benchmark spectrum. The solid black line represents the expected spectrum on Earth for neutrinos emitted with the SPL-NST benchmark spectrum (depicted by blue dashed line), in the absence of LIV. The crosses and shaded region carry the same meanings as those depicted in the middle panel of Fig. \ref{figure_1}.}
  \label{figure_3}
\end{figure}

Next, we consider neutrinos emitted at a specific redshift $z$, including propagation effects. The emitted energies lie within the intervals $[E_0, E_0+\Delta E]$ and $[E, E+\Delta E]$, where $\Delta E\ll E_0<E$. Let $N_0$ and $N$ denote the number of neutrinos within these two energy intervals, respectively. Then we have
\begin{equation}
    \frac{N}{N_0}=\frac{\Phi_z(E)\Delta E}{\Phi_z(E_0)\Delta E}=\frac{\Phi_z(E)}{\Phi_z(E_0)}=\left(\frac{E}{E_0}\right)^{-\alpha}.
    \label{Eq_B1}
\end{equation}
After propagation, the energy of the neutrinos is redshifted, causing the energy distribution intervals of these  $N_0$ and $N$ neutrinos to shift to $[E_0/(1+z), E_0/(1+z)+\Delta E/(1+z)]$ and $[E/(1+z), E/(1+z)+\Delta E/(1+z)]$, respectively. Let $\widetilde{\Phi}_z(z, E)$ denote the spectrum of neutrinos emitted at redshift $z$ after propagating to Earth. We then obtain
\begin{equation}
\frac{N}{N_0}=\frac{\widetilde{\Phi}_z\left(z,\frac{E}{1+z}\right)\frac{\Delta E}{1+z}}{\widetilde{\Phi}_z\left(z,\frac{E_0}{1+z}\right)\frac{\Delta E}{1+z}}=\frac{\widetilde{\Phi}_z\left(z,\frac{E}{1+z}\right)}{\widetilde{\Phi}_z\left(z,\frac{E_0}{1+z}\right)}.
\label{Eq_B2}
\end{equation}
From Eqs. \ref{Eq_B1} and \ref{Eq_B2}, it follows that
\begin{equation}
    \widetilde{\Phi}_z\left(z,\frac{E}{1+z}\right)=\widetilde{\Phi}_z\left(z,\frac{E_0}{1+z}\right)\left(\frac{E}{E_0}\right)^{-\alpha}\propto E^{-\alpha}.
\end{equation}
This implies that $\widetilde{\Phi}_z$ can be expressed as $\widetilde{\Phi}_z(z,E)\equiv f(z)E^{-\alpha}$, where $f(z)$ is a redshift-dependent function. Consequently, the overall spectrum of all neutrinos on Earth, after propagating from all redshifts, is given by
\begin{equation}
    \widetilde{\Phi}(E)\propto \int P(z)\widetilde{\Phi}_z(z,E)dz=E^{-\alpha}\int P(z)f(z)dz\propto E^{-\alpha}.
\end{equation}
Therefore, we have theoretically demonstrated that, in the absence of LIV, the expected spectrum on Earth retains the same power-law as the originally emitted spectrum.

We further verify this conclusion through Monte Carlo simulations. In Fig. \ref{figure_3}, the solid black line represents the expected spectrum of neutrinos on Earth, assuming an SPL-NST benchmark spectrum at emission and in the absence of LIV. It is obvious that the expected spectrum on Earth closely matches the emitted spectrum, represented by the blue dashed line. The right endpoint of the black curve falls below $10^4$ PeV , reflecting the maximum energy of emitted neutrinos in our simulation, which is reduced by cosmological redshift. The fluctuations at the high-energy end of the black curve result from statistical fluctuation, a consequence of the limited number of neutrinos in this energy range.

\end{appendix}
\end{document}